\documentclass[prb, superscriptaddress, floatfix,twocolumn,showpacs,amsfonts,amsmath,amssymb,final]{revtex4}

\usepackage{graphicx}

\usepackage{rotating}

\usepackage{amsmath}

\usepackage{amsfonts}

\usepackage{amssymb}

\usepackage[countmax]{subfloat}

\usepackage{dcolumn} 

\usepackage{bm} 

\usepackage{color}

\usepackage{float}

\usepackage{latexsym}

\begin{document}

\title{Fermi Surface Reconstruction by Dynamic Magnetic Fluctuations and Spin-Charge Separation Near an $O(3)$ Quantum Critical Point}


\author{Michael Holt}

\affiliation{School of Mathematics and Physics, University of Queensland, Brisbane, Queensland, 4072, Australia}


\author{Jaan Oitmaa}

\affiliation{School of Physics, University of New South Wales, Kensington 2052, Sydney NSW, Australia}

\author{Wei Chen}

\affiliation{Max-Planck-Institut f$\ddot{u}$r Festk$\ddot{o}$rperforschung, Heisenbergstrasse 1, D-70569 Stuttgart, Germany}

\author{Oleg P. Sushkov}

\affiliation{School of Physics, University of New South Wales, Kensington 2052, Sydney NSW, Australia}

\date{\rm\today}

\begin{abstract}

{Stimulated by the small/large Fermi surface controversy in the cuprates we consider a small number of holes injected into the bilayer antiferromagnet. The system has an $O(3)$ quantum critical  point (QCP) separating the magnetically ordered and the magnetically disordered phases. We demonstrate that nearly critical quantum magnetic  fluctuations can change the Fermi surface topology and also lead to spin charge separation (SCS) in two dimensions.  We demonstrate that in the physically interesting regime there is a magnetically driven Lifshitz point (LP) inside the magnetically disordered phase.  At the LP the topology of the hole Fermi surface is changed. The position of the LP, while being close to the position of the QCP is generally different. Dependent on the additional hole hopping integrals $t^{\prime}$ and $t^{\prime\prime}$, the LP can be located either in the magnetically ordered phase and/or in the magnetically disordered phase. We also demonstrate that in this regime the hole spin and charge necessarily separate when approaching  the QCP. The considered model sheds light on generic problems concerning the physics of the cuprates.}

\end{abstract}

\pacs{74.40.Kb, 74.72.Gh, 75.10.Jm, 75.50.Ee}





\maketitle

\section{Introduction}

\label{sec:intro}

It is well established that electron/hole diffraction from a static magnetic order in a conductor can influence the FS topology; this is FS reconstruction.  In addition it is also well established that spin and charge are separate in one-dimensional systems \cite{Tomonaga50, Luttinger63}. This results in two interesting but important generic questions; (i) Can {\it dynamic} magnetic fluctuations also drive a change in the FS topology? (ii) Is spin-charge separation possible in two-dimensions, and if so what is the meaning of the separation. In the present work we address these two generic problems and demonstrate that these two problems are remarkably related. These two important issues have recently attracted alot of attention due to their close tie to the high $T_{c}$ superconductivity in the cuprates. 

To address the two generic problems of interest we consider a small number of holes injected into the bilayer Heisenberg antiferromagnet where the interlayer coupling drives the strength of the magnetic fluctuations. The system has an $O(3)$ quantum critical point (QCP) separating the magnetically ordered and the magnetically disordered phases.  We show that indeed purely dynamic short range AF correlations in the absence of a static AF order can cause a LP in which the topology of the FS is changed. The position of the LP, while being close to the position of the QCP is generally different. Dependent on the additional hole hopping integrals $t'$ and $t''$, the LP can be located either in the magnetically ordered and/or in the magnetically disordered phase. Vojta and Becker \cite{Vojta99} also observed a LP in a similar model. However, in their study the LP was always in the AF ordered phase and therefore the central issue of the dynamic magnetic fluctuation driven LP has not yet been addressed.

We also demonstrate that the hole spin and charge necessarily separate when approaching the magnetic $O(3)$ QCP when the LP is in the magnetically disordered phase. The possibility of spin-charge separation (SCS) in 2D has been previously studied using the slave-boson method \cite{Putikka94, Chen94, Martins99}. However, the slave-boson method as applied to the $t-J$ model implies the SCS ad hoc \cite{WenLee}. This is not the case in the present work, therefore, the precise meaning of SCS in the present work is different from that of the slave-boson method.

Our interest in the two generic problems is motivated by the cuprates, which, in our opinion, manifest both the LP as well as SCS. It is therefore appropriate to explain the connection with the cuprates which we do below. A reader who is interested in these generic problems but not necessarily their application to the cuprates can go directly to Eq. (1) where we start the analysis.

One of the central issues in understanding high-$T_{c}$ superconductivity is whether it originates from a Fermi liquid or from a Mott insulator. While there is no consensus on the problem, there are experimental indications including angle-resolved photoemission spectroscopy (ARPES) that the transition from small to large Fermi surface  occurs in the hole doping range $0.1<x<0.15$ \cite{Marshall96, Zhou06, Yoshida07, Hossain08, Fournier10, He11, Yang11}. Fermi arcs \cite{Norman98} observed below this doping level have triggered various theoretical studies and models. However, recent ARPES data supports the existence of a shadow band that completes the arc into a pocket \cite{He11, Yang11}. The existence of hole pockets is in agreement with several calculations which considered dilute holes dressed by spin fluctuations, based on doping a Mott insulator \cite{Schmitt-Rink88, Kane89, Martinez91, Liu92, Ramsak92, Sushkov97}. On the other hand, a large Fermi surface as expected from a Fermi liquid approach is observed in ARPES studies in the optimally to overdoped cuprates. Naturally, this implies that there is at least one topological Lifshitz point  \cite{Lifshitz60} (LP)  in the doping range $0.1<x<0.15$, where the Fermi surface changes from small to large in contradiction to the Luttinger theorem. A phenomenological description based on Fermi liquid picture\cite{Yang06}, as well as dynamical mean-field theory calculations within Hubbard model\cite{Stanescu06} have been proposed to describe this LP. 

Magnetic quantum oscillation (MQO) in underdoped YBa$_{2}$Cu$_{3}$O$_{y}$ \cite{DoironLeyraud07} supports the small pocket scenario, in contrast to the large FS observed on the overdoped side \cite{Vignolle08}. The above observations also suggests the existence of a LP at which the topology of the FS changes. The MQO measurements were performed in very strong magnetic fields, up to 80T. It has been suggested \cite{Harrison09} that the field induces a static magnetic structure and the structure causes the small Fermi surface reconstruction. On the one hand, the MQO experiments suggest the small FS was observed up to 12\% doping \cite{Singleton10}, and it is unlikely that even an 80T field can generate a static AF order at such high doping. On the other hand the short range dynamic AF correlations always exist in the cuprates; this has been supported by recent RIXS measurements \cite{LeTacon11} which have remarkably demonstrated that such correlations are practically doping independent, from Mott insulator to optimal doping. Based on this data one can conjecture that the cuprates are always close to magnetic criticality. This motivates us to study if the LP can be driven by short range, purely dynamic AF correlations. We consider a bilayer model for the sake of performing a controlled calculation. However, we believe that conceptually our conclusions are equally applicable to both single and multi-layer cuprates.

Besides the possible topological transition, this particular doping range $0.1 < x < 0.15$ has also been associated with a quantum critical point (QCP) where the static magnetic ordering becomes fully dynamic. Below this doping range, neutron scattering indicates a commensurate or incommensurate magnetic ordering depending on the doping level as well as the coupling between the CuO$_2$ planes \cite{Yamada98, Hinkov04, Hinkov07, Hinkov08}. The static magnetic order vanishes within this doping range, as also confirmed by nuclear magnetic resonance (NMR) wipeout,\cite{Julien01,Kobayashi01} and recently confirmed by muon spin rotation ($\mu$SR) data \cite{Coneri10}.  The reduction of the static magnetic moment has been calculated within a low energy effective theory \cite{Milstein08}, and a QCP around $x\approx 0.11$ is predicted, close to the experimental value $x \approx 0.09$ observed in YBa$_{2}$Cu$_{3}$O$_{y}$.\cite{Stock08, Hinkov08, Haug10}. It is then intriguing to ask whether the topological LP is related to this magnetic QCP, especially how the change of magnetic excitations can influence the electronic properties. However, currently there is no precise measure to discern whether the topological transition takes place in the magnetically ordered or the fully dynamic phase.

Now we turn to the discussion of  SCS and its relation to magnetic quantum criticality. In La$_{2-x}$Sr$_x$CuO$_4$ the QCP is  smeared out because of disorder. However, in YBa$_{2}$Cu$_{3}$O$_{y}$ the QCP is located experimentally at doping  $x \approx 0.09$ ($y\approx 6.47$) \cite{Stock08, Hinkov08, Haug10}. SCS manifests itself in the cuprates as a phenomenon related to the magnetic QCP mentioned above. In the magnetically ordered phase, the pseudospin that marks the sublattice is a good quantum number for holes. The meaning of partial SCS is then the unusual way the pseudospin interacts with an external magnetic field \cite{Braz89, Ramaz08} and this is the meaning of the partial SCS in the magnetically ordered phase \cite{Milstein08}. On the other side of the QCP, the lack of long range order indicates that spin and charge are recombined, as expected from the Fermi liquid approach. However, it is unclear how SCS develops as one approaches the QCP from the magnetically disordered phase, or stating from another aspect, if there is a certain residual effect from SCS that affects the hole motion in the magnetically disordered phase. In addition, since the QCP and LP are located in close vincinity, is SCS related to or participate in the mechanism that causes FS reconstruction.  In the present work we analyse the process of SCS at the QCP. The model considered here has only commensurate magnetic ordering, so we put aside incommensurability in the cuprates. 

Our recent work \cite{Holt12} suggests that the two seemingly unrelated issues of FS reconstruction and SCS may originate from the same mechanism, namely nearly critical quantum fluctuations. By studying the bilayer AFM in the magnetically disordered phase, we show that dynamic magnetic fluctuations, driven by the ratio of the interlayer to the intralayer coupling $J_{\perp}/J$, has a dramatic effect on both the dispersion and spectral function of a single hole. The system under consideration has a $O(3)$ QCP at $(J_{\perp}/J)\approx 2.31$, close to which the dispersion has minima at ${\bf k}=(\pm\pi/2,\pm\pi/2)$. This results in small hole pockets similar to that in the cuprates at small doping. The pockets are formed due to strong in-plane AF correlations which diminish the nearest site hopping $t$. Upon increasing $J_{\perp}$ the in-plane AF correlations are reduced and the dispersion minima gradually shift toward ${\bf k}=(\pm\pi,\pm\pi)$ reaching this position at some value  $J_{\perp}>J_{\perp}^{LP}$. This is the position of the LP where the shape of the single hole dispersion changes. In-plane AF fluctuations are negligible at very large $J_{\perp}$, so the hole dispersion recovers the shape expected from a Fermi liquid approach, but the bandwidth is a factor of two reduced due to $J_{\perp}$. Note that we examine only the single hole case to demonstrate the mechanism, so the notion of a FS is ambiguious. Nevertheless, it is known that the SCBA typically converges at small doping up to $x\sim 0.1$ or so, and the rigid band approximation in most models is valid. Therefore we expect our results to be valid at small doping, where the FS changes from four small pockets centered around ${\bf k}=(\pi/2,\pi/2)$ at $J_{\perp}< J_{\perp}^{LP}$ to one small pocket centered at ${\bf k}=(\pi,\pi)$ at $J_{\perp}> J_{\perp}^{LP}$. Although it is unclear at present if this model can be realized in certain materials and if the driving parameter $J_{\perp}/J$ can be controlled externally, one clear prediction is that if an MQO experiment is available for such  a compound, then one would observe the MQO frequency changes by a factor of four across the LP. 

Taking these issues back to the cuprates, an intriguing argument for the relevance of our calculations comes from recent resonant inelastic x-ray scattering (RIXS) experiments \cite{LeTacon11}. RIXS indicates that magnetic fluctuations exist even in the overdoped regime, and is practically unrenormalized as in the undoped parent compound. Therefore purely dynamic magnetic fluctuations are very likely to be the driving force behind FS reconstruction, even though the LP in the cuprates takes place in a doping region that has no static magnetic order. However, one has to keep in mind that the magnetic fluctuations in the cuprates, regardless of the possible multilayer structure in a unit cell, are likely to be purely 2D, and the interlayer coupling is unlikely to be a crucial driving force for the magnetic fluctuations. Therefore, despite the appealing conclusions from the bilayer AFM, a theory that can address the 2D magnetic fluctuations in the cuprates will be much more challenging, and is yet to be formulated.     

The purpose of this extended version is to provide a detailed formalism for the results presented in Ref. \onlinecite{Holt12}, and to clarify several points that we omitted to address. We provide details for the bond operator formalism that describes the magnetic fluctuations in terms of triplon excitations. This mean field formalism yields a good agreement with Quantum Monte Carlo and series expansion results \cite{Sandvik94, Sandvik95, Zheng97}, and has the advantage that hole dynamics can be treated within a simple self-consistent Born approximation(SCBA). As mentioned in Ref. \onlinecite{Holt12}, the hole-triplon vertex has a discrepency with a previous study \cite{Brunger06}, so we provide detailed calculations to justify our vertex. To support the argument of magnetic fluctuation induced FS reconstruction, we have drawn evidence from series expansion calculations to justify our SCBA approach. In this long version we provide numerical results of the series expansion calculations and make a direct comparison with the SCBA calculations.  

The structure of the paper is the following. In Sec. \ref{sec:modelparam} we formulate the model and discuss the relevant set of parameters. In Sec. \ref{sec:bondoperatormft}, we present a brief overview of the bond operator formalism for the double layer Heisenberg model and reproduce the results of Refs. \cite{Matsushita99, Yu99} for spin excitations in the the magnetically disordered phase. In Sec. \ref{sec:holetriplonvertex} we derive the hole-triplon verticies for the quantum disordered phase, and compare them with previously known expressions. In Sec. \ref{sec:holegreen} we introduce the self-consistent Born approximation (SCBA) and calculate the hole Green's and spectral functions. Details of the numerical calculations are also presented. In Sec. \ref{sec:series} we outline the series expansion methods used in our study and in Sec. \ref{sec:results} discuss the results of our numerical calculations and comparison with dimer series expansions. In Sec. \ref{sec:scs} we address the issue of SCS in the vicinity of the QCP. Finally, in Sec. \ref{sec:conclusions} we present our conclusions.

\begin{figure}
 \begin{center}
     \includegraphics[trim=20mm 60mm 120mm 60mm, width=0.90\columnwidth,clip]{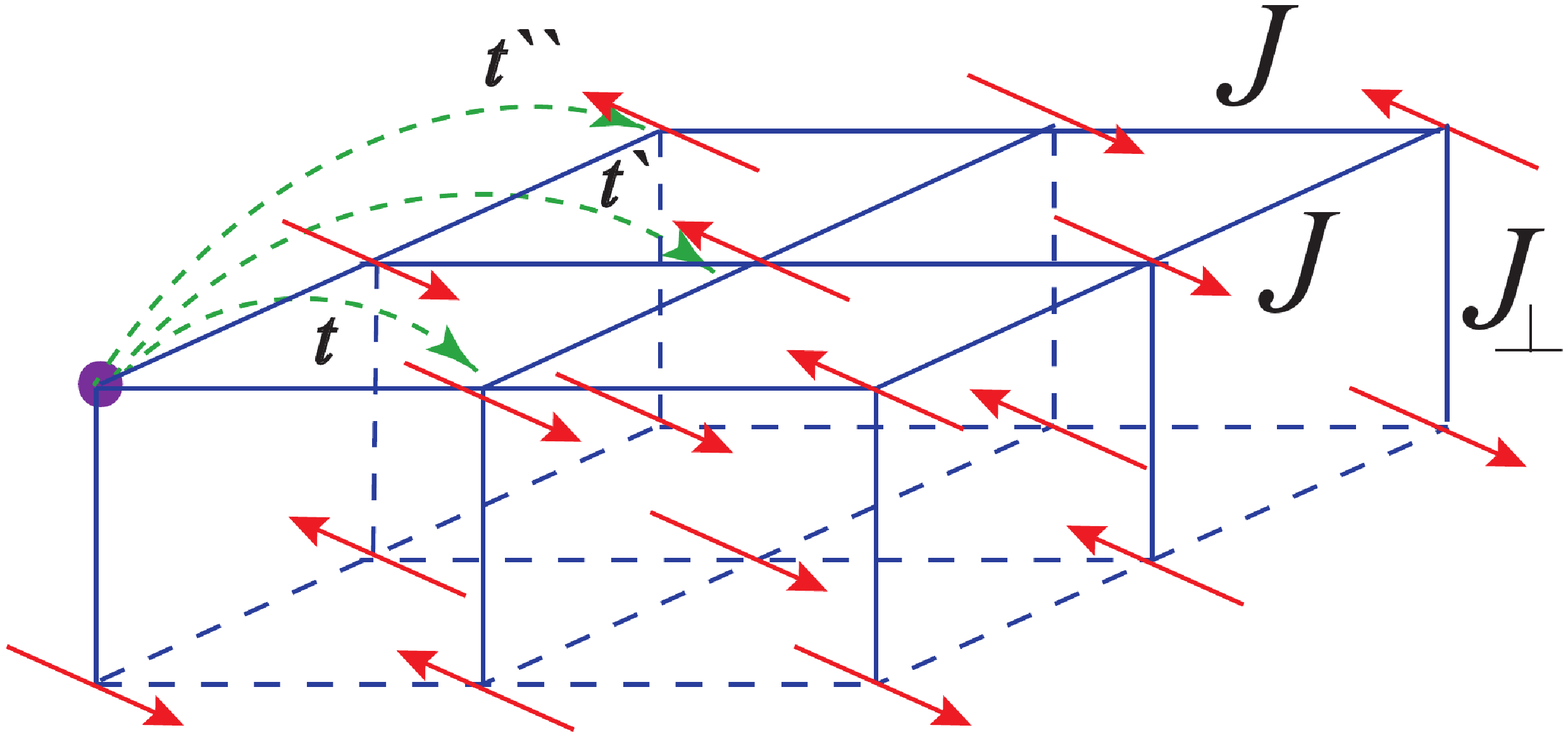}
    \end{center}
    \caption{(color online) Schematic diagram of the Heisenberg bilayer antiferromagnet doped with a single hole. We show the hole hopping integral parameters $t_{i,j}$, the antiferromagnetic exchange $J$ between spins on neighbouring sites as well as the interlayer antiferromagnetic exchange $J_{\perp}$.}
    \label{fig:schematic}
\end{figure}

\section{Model, Parameters}

\label{sec:modelparam}

We consider the $t-t'-t''-J-J_{\perp}$ model on a bilayer square lattice at zero temperature

\begin{eqnarray}
\label{eq:tJhamiltonian}
H &=& J\sum_{\langle i,j\rangle}({\bf S}_i^{(1)}\cdot {\bf S}_j^{(1)} + {\bf S}_i^{(2)}\cdot {\bf S}_j^{(2)}) + J_{\perp}\sum_{i}{\bf S}_i^{(1)}\cdot {\bf S}_i^{(2)} \nonumber\\
&-& \sum_{\langle i,j \rangle}t_{i,j}(c_{i1\sigma}^{\dagger} c_{j1\sigma} + c_{j1\sigma}^{\dagger} c_{i1\sigma})\nonumber \\
&=&H_{J,J_{\perp}}+H_{t,t^{\prime},t^{\prime\prime}}
\end{eqnarray}

\noindent where $c_{i\sigma}^{\dagger}$ is the creation operator of an electron with spin  $\sigma=\uparrow,\downarrow$  at site $i$ on the top plane, ${\bm S}_i^{(1)}=\frac{1}{2}c_{i\mu}^{\dagger}{\bm \sigma}_{\mu\nu}c_{i\nu}$, and $t_{i,j}=\left\{t, t^{\prime}, t^{\prime\prime}\right\}$ is the hopping integral between nearest, next-nearest, and next-next nearest neighbour sites respectively as shown in Fig. \ref{fig:schematic}. The superscripts (1), (2) in \eqref{eq:tJhamiltonian} indicate the layers. To compare parameters on a similar energy scale hereafter we set $J = 1$.  A no-double-occupancy constraint is imposed on the system. This means that at half filling (one electron per site) there are no mobile charge carriers and we have a Mott insulator. The $H_{J,J_{\perp}}$ term describes the antiferromagnetic coupling in each layer as well as between the two layers. It is well known that without holes (half-filling) the model has an O(3) magnetic QCP at $J_{\perp} = 2.525$ \cite{Sandvik94, Sandvik95} separating the magnetically disordered and the AF ordered phases. The hopping integrals $t,t',t''$ result in charge dynamics if a hole is injected into the system. Note that hopping is only allowed in the top plane, so a mobile hole can only be injected into the top layer. The longer range hopping integrals $t',t''$ are crucial as we will explain later. Note also that since we consider the zero temperature case, the magnetic ordering in the AF phase is consistent with the Mermin-Wagner theorem.

A similar extended $t-t^{\prime}-t^{\prime\prime}-J$ model has been widely applied to study the magnetic properties of the prototype bilayer cuprate YBa$_{2}$Cu$_{3}$O$_{6+y}$ in the regime $J_{\perp}/J \ll 1$, as indicated by fitting the magnon dispersion via neutron scattering data \cite{Tranquada89, Reznik96}. In the current study, we investigate the magnetically disordered phase in the opposite limit, $J_{\perp}/J> 1$, as addressed below.

In the present study we focus on small doping, $x \ll 1$, such that the magnetic fluctuations are not influenced by doping. Therefore we can assume that the magnetic fluctuations and the QCP is driven by the interlayer coupling $J_{\perp}$ alone. The holes  fill the rigid band formed by the magnetic quantum fluctuations. To address the problems formulated above it is sufficient to calculate the single hole Green's function. Certainly for sufficiently high doping the holes start to influence the magnetic fluctuations and hence the rigid band approach fails.  However, it is not necessary to go to such high doping to draw our conclusions.  Such an approach is only possible because the magnetic dynamics are driven by $J_{\perp}$ and are independent of the hole concentration. This is a significant simplification compared to the $t-J$/Hubbard model, where doping is the only ``handle''.

Although the current study does not intend to simulate the cuprates directly, we choose the parameters relevant to those in the cuprates such that the interplay of various mechanisms can be compared on a similar energy scale. The in-plane antiferromagnetic coupling in the cuprates as measured by two magnon Raman scattering \cite{Tokura90, Greven94} is $J = 125$ meV. As will be addressed in  later sections, we investigate several different hopping parameters and show the importance of $t^{\prime}$ and $t^{\prime\prime}$. To make a relevant connection to the cuprates, we choose one of the parameters close to that of YBa$_{2}$Cu$_{3}$O$_{7}$, which in the first principle study yields $t = 386$ meV, $t' = -105$ meV, $t'' = 86$ meV~\cite{Andersen95}.  In order to compare the interplay of various mechanisms on a similar energy scale we set $J = 1$. These values have also been applied to the analysis of photoemission in Sr$_2$CuO$_2$Cl$_2$,~\cite{Sushkov97} and yield a good agreement with the ARPES data. In the current approach we use the values provided in Table.\ \ref{tbl:LDAparam}. 

\begin{table}

\begin{tabular}{p{1.5cm}<{\centering}  p{1.5cm}<{\centering}  p{1.5cm}<{\centering} }

\hline \hline

    $t/J$ & $t'/J$ & $t''/J$ \\ \hline

   3.1 & -0.8 & 0.7 \\ \hline

   3.1 & 0.0 & 0.0 \\ \hline

   0.5 & 0.0 & 0.0 \\ \hline 

   0.5 & 0.0 & 0.1 \\ \hline \hline

    \end{tabular}	

\caption{Hopping integral parameters $t_{i,j}$ in units of the antiferromagnetic exchange $J$. Here we consider both {\it "strong"} ($t/J=3.1$) and {\it "weak"} ($t/J=0.5$) coupling.}

\label{tbl:LDAparam}

\end{table}

The motivation of using these parameters is that they cover both {\it "strong"} and {\it "weak"} coupling regimes. Here we denote {\it "strong"} with large $t$ and {\it "weak"} with small $t$. The importance of the {\it "strong"} coupling regime is stressed by its  relevance to the cuprates. The {\it "weak"} coupling regime is particularly useful as comparison with series expansions is possible. In both these limits we consider the "pure" $t-J$ model which corresponds to $t' = t'' = 0$ and also a case with nonzero $t',t''$.

\section{Bond-Operator Mean-Field Theory}

\label{sec:bondoperatormft}

The magnetic excitations in the magnetically disordered phase are triplons. To describe the triplons we employ the spin-bond operator mean field technique. This approach has been previously applied to quantum disordered systems such as bilayer antiferromagnets, spin chains, spin ladders,  Kondo insulators etc.\cite{Jurecka00, Eder98, Saito03, Vojta99, Matsushita99, Yu99}.  It is known~\cite{Matsushita99, Yu99} that  this simple technique gives a QCP at $(J_{\perp}/J)_{c} \approx 2.31$, which is close to the exact value $(J_{\perp}/J)_{c} = 2.525$ known from Quantum Monte Carlo calculations \cite{Sandvik94, Sandvik95}, series expansions \cite{Zheng97}, and more involved analytical calculations \cite{Kotov98}. 

In the present work we consider only the magnetically disordered phase since it proved to be sufficient to locate the LP. In principle one could extend the present study to the magnetically ordered phase. The bond-operator mean-field technique has been considered for the ordered phase of the bilayer antiferromagnet by both Normand \cite{Normand97} and Vojta \cite{Vojta99}. This could be considered in a future study. However, this is beyond the scope of the present work.

All necessary equations describing the triplon dynamics have been derived previously \cite{Yu99}. We present here a brief overview of the technique in application to the bilayer antiferromagnet. Our aim is to establish a reliable  description and to determine parameters which will be subsequently used in the SCBA calculations of a dressed hole, as addressed in Sec. \ref{sec:holegreen}. One can certainly employ a more accurate  Brueckner technique \cite{Kotov98}. However, this technique  is more involved while the bond operator mean field approach has sufficient accuracy for our purposes and we chose it for simplicity.

The bond-operator representation describes the system in a base of pairs of coupled spins on a rung, which can either be in a singlet or triplet (triplon)  state:
\begin{eqnarray}
\label{eq:bondoperator}
\mid s\rangle_i &=& s_i^\dagger \mid 0\rangle_i = \frac{1}{\sqrt{2}}(\mid \uparrow \downarrow\rangle_i - \mid\downarrow \uparrow\rangle_i)\nonumber\\
\mid t_x\rangle_i &=& t_{ix}^\dagger \mid0 \rangle_i = \frac{-1}{\sqrt{2}}(\mid \uparrow \uparrow\rangle_i - \mid \downarrow \downarrow \rangle_i)\nonumber\\
\mid t_y \rangle_i &=& t_{iy}^\dagger\mid 0>_i = \frac{i}{\sqrt{2}}(\mid\uparrow \uparrow\rangle_i + \mid\downarrow \downarrow\rangle_i)\nonumber\\
\mid t_z\rangle_i &=& t_{iz}^\dagger\mid 0>_i = \frac{1}{\sqrt{2}}(\mid\uparrow \downarrow\rangle_i + \mid\downarrow \uparrow\rangle_i)
\end{eqnarray}
where the four types of bosons obey the bosonic commutation relations. To restrict the physical states to either singlet or triplet, the above operators are subjected to the constraint,
\begin{equation}
\label{eq:constraint}
s_i^\dagger s_i + \sum_{i,\alpha}t_{i\alpha}^\dagger t_{i\alpha} = 1
\end{equation}
\noindent In terms of these bosons, the spin operators in each layer ${\bf S}_i^{(1)}$ and ${\bf S}_i^{(2)}$ can be expressed as
\begin{equation}
\label{eq:spins}
S_{i\alpha}^{(1,2)} = \frac{1}{2}(\pm s_i^\dagger t_{i\alpha} \pm t_{i\alpha}^\dagger s_i - i\sum_{\beta \gamma}\epsilon_{\alpha \beta \gamma}t_{i\beta}^\dagger t_{i\gamma})
\end{equation} 
\noindent where $\alpha$ represents respectively the components along the $x$, $y$, and $z$ axes. Substituting the bond-operator representation of spins defined in Eq. \eqref{eq:spins} into the $H_{J,J^{\prime}}$ in Eq. \eqref{eq:tJhamiltonian} we obtain 
\begin{eqnarray}
\label{eq:bondoperatorhamiltonian}
H_{J,J_{\perp}} &=& H_1 + H_2 + H_3 + H_4\nonumber\\
H_1 &=& J_{\perp}\sum_{i}(-\frac{3}{4}s_i^\dagger s_i + \frac{1}{4}t_{i\alpha}^\dagger t_{i\alpha}) 
\nonumber \\
H_2 &=& \frac{J}{2}\sum_{\langle i,j\rangle}(s_i^\dagger s_j^\dagger t_{i\alpha} t_{j\alpha} + s_i^\dagger s_j t_{i\alpha} t_{j\alpha }^\dagger + H.c.)\nonumber\\
H_3 &=& \frac{J}{2}\sum_{\langle i,j \rangle}(i\epsilon_{\alpha \beta \gamma} t_{j\alpha}^\dagger t_{i\beta}^\dagger t_{i\gamma} s_j + i\epsilon_{\alpha \beta \gamma} t_{j\alpha} t_{i\beta} t_{i\gamma}^\dagger s_j^\dagger)\nonumber\\
H_4 &=& \frac{J}{2}\sum_{\langle i,j \rangle}(1 - \delta_{\alpha\beta})(t_{i\alpha}^{\dagger} t_{j\beta}^{\dagger} t_{i\beta} t_{j\alpha} -  t_{i\alpha}^{\dagger} t_{j\alpha}^{\dagger} t_{i\beta} t_{j\beta})\nonumber \\
&&
\end{eqnarray}

The main issue is how to account for the hard core constraint (\ref{eq:constraint}). In principle this can be done via an infinite on-site repulsion of triplon excitations, however this technique is quite involved \cite{Kotov98}. Our aim is the hole dispersion, therefore for magnetic excitations the simplest possible technique which reproduces the magnetic QCP is sufficient. This is why we employ the simple mean field approach~\cite{Matsushita99,Yu99} that accounts the constraint (\ref{eq:constraint}) via a Lagrange multiplier $\mu$ in the Hamiltonian
\begin{eqnarray}
\label{h11}
H_1 &\to& H_1-\mu \sum_{i} (s_i^\dagger s_i + t_{i\alpha}^\dagger t_{i\alpha} - 1)\\
&=&J_{\perp}\sum_{i}(-\frac{3}{4}s_i^\dagger s_i + \frac{1}{4}t_{i\alpha}^\dagger t_{i\alpha})
-\mu \sum_{i} (s_i^\dagger s_i + t_{i\alpha}^\dagger t_{i\alpha} - 1)\ .\nonumber
\end{eqnarray}
\noindent Further analysis is straightforward. We neglect the cubic and quartic Hamiltonians $H_3$ and $H_4$; replace singlet operators by numbers, $\langle s_{i}^{\dagger} \rangle = \langle s_{i} \rangle = \bar{s}$ (Bose-Einstein condensation of spin singlets); and finally diagonalize the quadratic in $t$ Hamiltonian by performing the usual Fourier and Bogoliubov transformations
\begin{equation}
\label{eq:Fourier}
t_{i\alpha} = \sqrt{\frac{1}{N}}\sum_{q}e^{iq\cdot r_{i}}t_{q\alpha} \  \  a_{i\sigma} = \sqrt{\frac{1}{N}}\sum_{k}e^{ik\cdot r_{i}}a_{k\sigma}  
\end{equation}
\begin{equation}
\label{eq:Bogoliubov}
\beta_{q\alpha} = u_{q} t_{q\alpha} - v_{q}t_{-q\alpha}^{\dagger} \   \ \beta_{q\alpha}^{\dagger} = u_{q} t_{q\alpha}^{\dagger} - v_{q}t_{-q\alpha}
\end{equation}
\noindent Here $N$ is the number of dimers; the diagonalized Hamiltonian reads
\begin{eqnarray}
\label{eq:diagonalizedmfa}
H_{MFA}(\mu, \bar{s}) &=& N\bigg (-\frac{3J_{\perp} s^2}{4} - \mu \bar{s}^2 + \mu \bigg ) + \frac{3}{2}\sum_{q}(\Omega_{q} - A_{q})\nonumber\\
&+& \sum_{\bf q}\Omega_{q}\beta_{q\alpha}^\dagger \beta_{q\alpha}
\end{eqnarray}
\noindent where
\begin{eqnarray}
\label{eq:coefficients}
\Omega_{q} &=& \sqrt{A_q^2 - 4 B_q^2}\nonumber\\
A_q &=& \frac{J_{\perp}}{4} - \mu + J\bar{s}^2(cos(q_x) + cos(q_y))\nonumber\\
B_q &=& \frac{J\bar{s}^2}{2}(cos(q_x) + cos(q_y))
\end{eqnarray}

\noindent Here the lattice spacing has been taken to be unity. The Bogoliubov coefficients $u_q$ and $v_q$  are given by
\begin{eqnarray}
\label{eq:bogoliubovcoeff}
u_q &=& \sqrt{\frac{A_q}{2\Omega_{q}} + \frac{1}{2}}\nonumber\\
v_q &=& -sign(B_{q})\sqrt{\frac{A_q}{2\Omega_{q}} - \frac{1}{2}} 
\end{eqnarray}
\noindent The parameters $\mu$ and $\bar{s}$ are determined by the saddle point equations:
\begin{equation}
\label{eq:saddle}
\bigg \langle \frac{\partial H_{MFA}}{\partial \mu} \bigg \rangle = 0 \ \ \bigg \langle \frac{\partial H_{MFA}}{\partial \bar{s}} \bigg \rangle = 0
\end{equation}
\noindent Following  Ref. \onlinecite{Matsushita99}, it is convenient to introduce the dimensionless parameter $d$
\begin{equation}
\label{eq:dimparamd}
d =\frac{2J\bar{s}^2}{\frac{J_{\perp}}{4} - \mu } 
\end{equation}
\noindent which results in the following self-consistency equations
\begin{eqnarray}
\label{eq:selfconsistentparamd}
d &=&  \frac{J}{J_{\perp}}\bigg (5 - \frac{3}{N}\sum_{q}\frac{1}{\sqrt{1+2d\gamma_{q}}} \bigg)\nonumber\\
\bar{s}^2 &=& \frac{5}{2} - \frac{3}{2N}\sum_{q}\frac{1+d\gamma_{q}}{\sqrt{1+2d\gamma_{q}}}\nonumber\\
\mu &=& -\frac{3J_{\perp}}{4}+\frac{3J}{N}\sum_{q}\frac{\gamma_{q}}{\sqrt{1+2d\gamma_{q}}}
\end{eqnarray}
\noindent Once the parameter $d$ is determined from the first  equation, the values of $\bar{s}$ and $\mu$ can be calculated using the second/third relations. These parameters Eq. \eqref{eq:selfconsistentparamd} are then used to determine the excitation spectrum 
of the system
\begin{equation}
\label{eq:spectrum}
\Omega_{q} =  \bigg (\frac{J_{\perp}}{4} - \mu \bigg ) \bigg [1 + 2d\gamma_{q} \bigg ]^{1/2}
\end{equation}

\begin{figure}
 \begin{center}
     \includegraphics[trim= 0mm 0mm 0mm 0mm, width=0.95\columnwidth,clip]{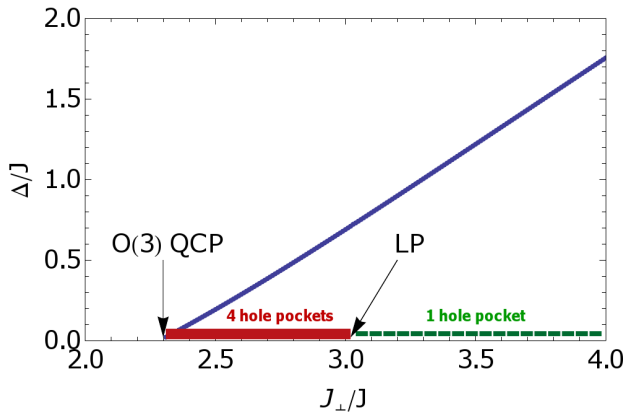}
    \end{center}
    \caption{(color online) Triplon gap $\Delta$ at $(\pi, \pi)$ as a function of $J_{\perp}/J$. The solid and dashed areas along the $J_{\perp}/J$-axis indicate the different topologies of the hole dispersion for the parameters $t=3.1$, $t^{\prime}=-0.8$, $t^{\prime\prime}=0.7$, same as in Fig. \ref{fig:spectralfermit310tp080tpp070}(b)-(d). Here the "4 hole pockets" region describes the four disconnected hole pockets with minima at ${\bf k} = (\pm \pi/2, \pm \pi/2)$ while the "1 hole pocket" region describes a single hole pocket with minimum at ${\bf k} = (\pi, \pi)$.}.
    \label{fig:spingapvelocity}
\end{figure}

As a characteristic parameter of the magnetically disordered ground state, we will be most interested in the value of the spin gap, i.e. the minimum energy gap of the triplon excitations, which is given by
\begin{equation}
\label{eq:spingap}
\Delta_{(\pi, \pi)} =\bigg (\frac{J_{\perp}}{4} - \mu \bigg ) \bigg [1 - 2d \bigg ]^{1/2}
\end{equation}
\noindent The results of numerical calculations of the spin gap is presented in Fig. \ref{fig:spingapvelocity}. As $J_{\perp}$ decreases from infinity, the triplet gap at the wave vector $q = (\pi,\pi)$ decreases, and the gap vanishes linearly at $J_{\perp}/J = (J_{\perp}/J)_{c}$, signalling the magnetic instability of the dimer phase.  The critical value can be obtained by setting $d = 1/2$ in Eq. \eqref{eq:selfconsistentparamd}, which yields $ (J_{\perp}/J)_{c}\approx 2.3$, in reasonable agreement with accurate 
Quantum Monte Carlo and series expansions result, $ (J_{\perp}/J)_{c}\approx 2.5$.~\cite{Sandvik94,Sandvik95,Zheng97}
Typical values of the parameters are presented in Table \ \ref{tab:meanfield}.

\begin{table}

\begin{center}

\begin{tabular}{c c c c c}

\hline \hline

$J_{\perp}/J$ & $\bar{s}$ & $\mu/J$ &$\left(\mu +\frac{3}{4}J_{\perp}\right)/J$ &$\Delta_{(\pi, \pi)}/J$ \\

\hline

2.31 & 0.906 & -2.703 &-0.971 &0.021 \\

\hline

2.50 & 0.918 & -2.756 &-0.881&0.192 \\

\hline

2.70 & 0.929 & -2.821 &-0.796 &0.384 \\

\hline

3.00 & 0.943 & -2.938 &-0.688 &0.689 \\

\hline

3.50 & 0.960 & -3.178 &-0.553 &1.218 \\

\hline

4.00 & 0.971 & -3.459 &-0.459 &1.754\\

\hline \hline

\end{tabular}

\caption{\label{tab:meanfield} Self consistent solution of the mean field parameters as a function of $J_{\perp}$ at zero temperature. Calculations were performed by self consistently solving Eq. \eqref{eq:selfconsistentparamd} and substituting the resulting parameters in Eq. \eqref{eq:spingap}.}

\end{center}

\end{table}

As one should expect the simple mean field approach described in this section does not give the correct critical index for the gap, see Fig. \ref{fig:spingapvelocity}, however, for calculation of the hole Green's function the obtained accuracy is sufficient. One can think that the value of the ``chemical potential'' $\mu$ is a measure of correlations. Values of $\mu$ presented in Table \ \ref{tab:meanfield} are surprisingly large. However, looking at Eq.(\ref{h11}) we see that $\frac{3}{4}J_{\perp}$ is just a redefinition of the ground state energy. Therefore, the true measure of correlations is $\mu+\frac{3}{4}J_{\perp}$. This quantity also  presented in Table  \ \ref{tab:meanfield} is not too large and approaches zero when $J_{\perp}\to \infty$.

\section{Hole-Triplon Vertex}

\label{sec:holetriplonvertex}

At half filling (one electron per site) the model under consideration is equivalent to a Heisenberg model; specifically, it represents a Mott insulator with no long range antiferromagnetic order since we are considering the magnetically disordered  phase. In the triplon technique used in Section III the magnetically disordered state is built on the perturbative ground state consisting of isolated spin singlets (Bose condensation of spin singlets). Triplon virtual excitations accounted via Bogoliubov's transformation gives the physical ground state in terms of the perturbative ground state. Since we will be using a diagrammatic technique we need to define a hole in terms of the same perturbative ground state. Hence we define the hole  creation operator with spin $\sigma$ by its action on the spin singlet bond $i$ by $a_{i,\sigma}^{\dagger}$
\begin{eqnarray}
\label{eq:holon}
&&a_{i\uparrow}^{\dagger} \mid s\rangle = \mid \circ \uparrow \rangle
=c_{i2\uparrow}^{\dag}\mid 0 \rangle\nonumber\\
&&a_{i\downarrow}^{\dagger} \mid s\rangle = 
\mid \circ \downarrow \rangle
=c_{i2\downarrow}^{\dag}\mid 0 \rangle \ ,
\end{eqnarray} 
\noindent where $\mid 0 \rangle$ is vacuum. Note that the hole is created only at the upper plane as we have discussed in Section II. The electron annihilation operator in the upper plane can be expressed in terms of $a_{i\uparrow}^{\dagger}$, see e.g. Ref.~\cite{Jurecka00}
\begin{eqnarray}
\label{eq:electronbondoperator}
c_{i1\sigma} = \frac{1}{\sqrt{2}}
\bigg [a_{i,\bar{\sigma}}^{\dagger}(p_{\sigma}s_i + t_{ix}) 
+ a_{i,\sigma}^{\dagger}(p_{\bar{\sigma}}t_{ix} + it_{iy}) \bigg]\ ,
\end{eqnarray}
\noindent where $p_{\sigma} = +(-)$, $\bar{\sigma} =\downarrow(\uparrow)$ for $\sigma = \uparrow(\downarrow)$. We also have to modify the hard core constraint (\ref{eq:constraint})
\begin{equation}
\label{eq:hardcoreconstraint}
s_{i}^{\dagger} s_{i} + \sum_{\alpha}t_{i\alpha}^{\dagger} t_{i\alpha} + \sum_{\sigma} a_{i\sigma}^{\dagger} a_{i\sigma} = 1 \ ,
\end{equation}
\noindent however, this modification is not important for the single hole problem as well as it is not important for the low density hole problem. Note that the operators $a_{i\sigma}^{\dagger}$ are required to obey fermionic anticommutation relations. This is important even in the single hole problem, since due to statistics, the sign of the hole dispersion is opposite to that of an electron.

Substitution of (\ref{eq:electronbondoperator}) in $H_{t,t^{\prime},t^{\prime\prime}}$ defined in (\ref{eq:tJhamiltonian}) gives the Hamiltonian expressed in terms of hole and triplon operators
\begin{eqnarray}
\label{eq:holehoppingfull}
&&H_{t,t^{\prime},t^{\prime\prime}}= \sum_{\langle i,j \rangle \sigma}\frac{t_{i,j}\bar{s}^2}{2}(a_{i, \sigma}^{\dagger}a_{j,\sigma} + a_{j, \sigma}^{\dagger}a_{i,\sigma})\nonumber\\ 
&+& \sum_{\langle i,j \rangle \sigma}\frac{t_{i,j}}{2}((t_{i\alpha}^{\dagger}t_{j\alpha}a_{j,\sigma}^{\dagger}a_{i,\sigma} + t_{j\alpha}^{\dagger}t_{i\alpha}a_{i,\sigma}^{\dagger}a_{j,\sigma}) \nonumber\\
&-&\sum_{\langle i,j \rangle \sigma}(t_{j\alpha}^{\dagger} \cdot [t_{i,j}\bar{s} \vec{T}_{i,j} + \frac{J\bar{s}}{2} \vec{T}_{i,i}] + t_{i\alpha}^{\dagger} \cdot [t_{i,j}\bar{s} \vec{T}_{j,i} + \frac{J\bar{s}}{2}  \vec{T}_{j,j}]\nonumber\\
&+& t_{j\alpha} \cdot [t_{i,j}\bar{s} \vec{T}_{j,i} + \frac{J\bar{s}}{2} \vec{T}_{i,i}] + t_{i\alpha} \cdot [t_{i,j}\bar{s} \vec{T}_{i,j} + \frac{J\bar{s}}{2}  \vec{T}_{j,j}])\nonumber\\
&-& \sum_{\langle i,j \rangle}it_{i,j}(\vec{T}_{j,i}\cdot(t_{i\alpha}^{\dagger} \times t_{j\alpha}) + \vec{T}_{i,j}\cdot(t_{j\alpha}^{\dagger} \times t_{i\alpha}))\nonumber\\ 
&-& \frac{J}{2}\sum_{\langle i,j \rangle}i(\vec{T}_{i,i}\cdot(t_{j\alpha}^{\dagger}\times t_{j\alpha}) +\vec{T}_{j,j}\cdot(t_{i\alpha}^{\dagger}\times t_{i\alpha})) \ .
\end{eqnarray}
\noindent Here 
\begin{equation}
\label{eq:interaction}
\vec{T}_{i,j} =\frac{1}{2}\sum_{\sigma, \bar{\sigma}} a_{i\sigma}^{\dagger}\vec{\tau}_{\sigma \bar{\sigma}}a_{j\bar{\sigma}} \ 
\end{equation}
\noindent where ${\vec {\tau}}$ is the Pauli matrix. Eq. \eqref{eq:holehoppingfull} contains hopping without changing the spin background (direct hopping), as well as spin-fluctuation assisted hopping, and exchange processes where the hole remains on its dimer. Considering the direct hopping terms in Eq. \eqref{eq:holehoppingfull} one obtains the bare hole dispersion
\begin{eqnarray}
\label{eq:fullholedispersion}
\epsilon_{k}^{(0)} = 2t\bar{s}^2\gamma_{k}
 + 2t'\bar{s}^2\gamma_{k}^{'}
+  2t''\bar{s}^2\gamma_{k}^{''} \ ,
\end{eqnarray}  
\noindent where
\begin{eqnarray}
\label{eq:lattice}
\gamma_{k} &=& \frac{1}{2}(cos(k_x) + cos(k_y))\nonumber\\
\gamma_{k}^{'} &=&  cos(k_x)cos(k_y)\nonumber\\
\gamma_{k}^{''} &=&  \frac{1}{2}(cos(2k_x) + cos(2k_y))
\end{eqnarray}

The most important interaction is the hole-triplon vertex described by the effective Hamiltonian
\begin{equation}
\label{eq:spinholefull}
\sum_{k q}\sum_{\sigma \bar{\sigma}}g_{k,q}\beta_{q\alpha}\cdot (a_{k+q\sigma}^{\dagger}\tau^{\alpha}_{\sigma \bar{\sigma}} a_{k\bar{\sigma}}) + H.c.
\end{equation}
\noindent The interaction is due to the third and fourth lines in Eq.(\ref{eq:holehoppingfull}). Taking matrix elements of these lines
we find the verticies
\begin{eqnarray}
\label{eq:coupling}
g_{k,q} &=& g^{a}_{k,q}+g^{b}_{k,q}+g^{c}_{k,q}+g^{d}_{k,q}\\
g^{a}_{k,q} &=& -\frac{2t\bar{s}}{\sqrt{N}}(\gamma_{k}u_{q} + \gamma_{k+q}v_{q})
\nonumber\\
g^{b}_{k,q} &=& -\frac{2t'\bar{s}}{\sqrt{N}}(\gamma'_{k}u_{q} + \gamma'_{k+q}v_{q})
\nonumber\\
g^{c}_{k,q} &=& -\frac{2t''\bar{s}}{\sqrt{N}}(\gamma''_{k}u_{q} + \gamma''_{k+q}v_{q})
\nonumber\\
g^{d}_{k,q} &=& -\frac{J\bar{s}}{\sqrt{N}}\gamma_{q}(u_{q} + v_{q})
\nonumber
\end{eqnarray}
\noindent Note, that on approaching the QCP the verticies diverge at ${\bf q}\to (\pi,\pi)$. The verticies (\ref{eq:coupling}) differ both in the coefficients and in the kinematic structure from the verticies obtained in Ref. \onlinecite{Brunger06}. We believe that their verticies~\cite{Brunger06} are not correct.

The Hamiltonian (\ref{eq:holehoppingfull}) generates hole-double-triplon verticies as well. They are determined by the second, 5th and 6th lines of (\ref{eq:holehoppingfull}). However, these verticies are due to quantum fluctuations of triplons and therefore they are very small. We neglect such verticies.

Thus the effective Hamiltonian used in the next section for the calculation of the hole Green's function is of the following form
\begin{eqnarray}
\label{eq:fullHamiltonian}
H &=& \sum_{q} \Omega_{q} \beta_{q\alpha}^{\dagger} \beta_{q\alpha} + \sum_{k,\sigma} \epsilon_{\bf k}^{(0)} 
a_{k\sigma}^{\dagger} a_{k\sigma}  \\
&&+\sum_{k,q,\sigma,\bar{\sigma}}\left\{ g_{k,q} \beta_{q\alpha} \cdot (a_{k+q\sigma}^{\dagger} \tau_{\sigma\bar{\sigma}} a_{k\bar{\sigma}}) + H.c.\right\} \nonumber 
\end{eqnarray}

\section{Hole Green's function}

\label{sec:holegreen}

\begin{figure}[h]
 \begin{center}
     \includegraphics[width=0.95\columnwidth,clip]{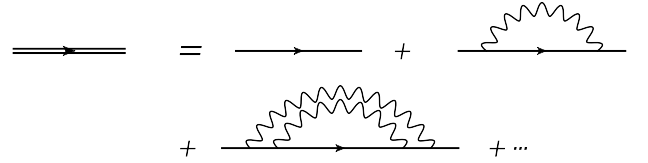}
    \end{center}
    \caption{Dysons' equation in the the self-consistent Born approximation.}
    \label{fig:SCBArainbowdiagrams}
\end{figure}

\begin{figure}[h]
 \begin{center}
     \includegraphics[width=0.45\columnwidth,clip]{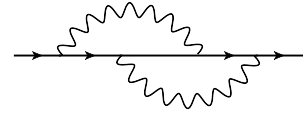}
    \end{center}
    \caption{Vertex corrections not included in the SCBA.}
    \label{fig:vertexcorrections}
\end{figure}

The retarded Green functions for a hole is defined in the standard way
\begin{equation}
\label{eq:pseudofermiongreen}
G_{\sigma}({\bf k},\omega) = -i\int_{0}^{\infty}\langle gs \mid  a_{{\bf k\sigma}}(t) a_{{\bf k}\sigma}^{\dagger}(0)  \mid gs \rangle e^{i\omega t} dt
\end{equation}
\noindent where $\mid gs \rangle$ is the ground state of the system. To describe the hole dressed by triplons we use the self-consistent Born approximation (SCBA) which disregards vertex corrections  in Dysons' equation for the hole Green's function; it amounts to an infinite order summation of noncrossing triplon loops (rainbow diagrams) in the hole self-energy,  as presented diagramatically in Fig.~\ref{fig:SCBArainbowdiagrams}. Vertex corrections shown in Fig.~\ref{fig:vertexcorrections} are not accounted for in the SCBA. This approximation gives the leading term in the $1/N$ expansion for the $O(N)$ group \cite{Sushkov00}

This approximation has been widely used to study hole dynamics in the AF background  \cite{Schmitt-Rink88, Kane89, Martinez91, Liu92, Ramsak92, Sushkov97}. In the case of the AF background the vertex correction, Fig. \ref{fig:vertexcorrections}, is equal to zero due to kinematic constraints \cite{Martinez91, Liu92, Sushkov94} and hence the SCBA is very accurate. In the present case of the magnetically disordered background the single loop vertex correction is  nonzero. However, the correction is suppressed by the parameter $1/N$, where $N = 3$ is the number of the triplon components \cite{Sushkov00}. To confirm the accuracy of the SCBA we compare results with that of numerically exact dimer series expansions. Note that here we are working in terms of the true spin of the hole, while in the case of the AF background one has to work in terms of pseudospin.

The hole dispersion in the magnetically ordered phase is strongly constrained by symmetry stemming from the magnetic ordering \cite{Chen11}. However, here we consider the magnetically disordered phase and hence the dispersion is free of such a constraint. The hole dispersion is purely determined by the bare hole hopping and the underlying spin dynamics, similar to the cuprates at high doping levels where the magnetic ordering becomes fully dynamic.

The zero approximation Green's function is
\begin{equation}
\label{eq:barepseudofermion}
G_{\sigma}^{0}({\bf k},\omega) = \frac{1}{\omega - \epsilon_{\bf k}^{(0)} + i0} \ ,
\end{equation}
\noindent where $\epsilon_{\bf k}^{(0)}$ is bare dispersion (\ref{eq:fullholedispersion}). Summation of diagrams  Fig.~\ref{fig:SCBArainbowdiagrams} leads to the following Dyson's equation
\begin{equation}
\label{eq:fullholegreen}
G_{\sigma}({\bf k},\omega) = \bigg (\omega - \epsilon_{\bf k}^{(0)}  
- 3\sum_{q} g_{{\bf k - q,q}}^{2} G_{\sigma}({\bf k - q},\omega - \Omega_{q})  + i\eta \bigg )^{-1} \ 
\end{equation}
\noindent where the factor $3$ comes from three different polarizations of the intermediate triplon. Similar to the argument in Ref. \onlinecite{Sushkov97}, at infinite $\omega$ the Green's function satisfies $G_{\sigma} = G_{\sigma}^{0}$, therefore the sum rule
\begin{equation}
\label{eq:sumrule}
-\frac{1}{\pi} Im \int_{-\infty}^{\infty} G_{\sigma}({\bf k},\omega) d\omega = 1
\end{equation}
\noindent is recovered.

We solve Dyson's equation, Eq. \eqref{eq:fullholegreen}, numerically on a $128\times 128$ cluster with energy resolution $\Delta\omega=0.02J$ and an artificial broadening $\eta=0.02J$. The spectral function of a dressed hole is  calculated by
\begin{equation}
\label{eq:spectraldensity}
A({\bf k},\omega) = -\frac{1}{\pi}Im[G({\bf k},\omega)]
\end{equation}

\section{Dimer Series Expansions}

\label{sec:series}

To check the accuracy of the SCBA approach we have also used series expansion methods \cite{Oitmaa99, Oitmaa06} to compute the hole excitation energies. The approach we have used is to treat the $J_{\perp}$ terms exactly ('dimer expansion') and the other terms as perturbations. A linked-cluster expansion is used, wherein an effective Hamiltonian is derived for the 1-hole sector of each cluster. This yields a set of transition amplitudes for hopping of the hole through different lattice vectors, which can be summed to yield series for excitation energies for any wavevector $\bf k$. The series are then analysed by standard Pade approximant methods. We refer the reader to Ref. \onlinecite{Oitmaa06} for further details of the method. In the present calculation, due to the presence of the further neighbor hopping terms $t'$,$t''$ the number of clusters grows rather rapidly. We have carried out the calculations to order 7, involving a total of 44312 distinct clusters with up to 7 dimers.

Despite the rather short series, convergence is good provided the $t$-parameters are not too large. By its nature, the series method will be accurate at small $t$-values whereas the SCBA is expected to be most accurate for large $t$.

\section{Numerical Results}

\label{sec:results}

First of all we remind the well known result~\cite{Schmitt-Rink88, Kane89, Martinez91, Liu92, Ramsak92, Sushkov97} that in the magnetically ordered phase at $J_{\perp}=0$ the hole dispersion minimum is located at  ${\bf k}=(\pm\pi/2,\pm\pi/2)$. On the other hand at $J_{\perp} \gg J$ the hole dispersion is given by Eq.(\ref{eq:fullholedispersion}) with ${\bar s}=1$. We always take $t > t',t''$, therefore the minimum of the dispersion is located at ${\bf k}=(\pm\pi,\pm\pi)$. This implies that there is at least one Lifshitz point (LP) between $J_{\perp}=0$ and $J_{\perp} =\infty$. Below we present and discuss our results and locate the LP. 

\begin{figure}
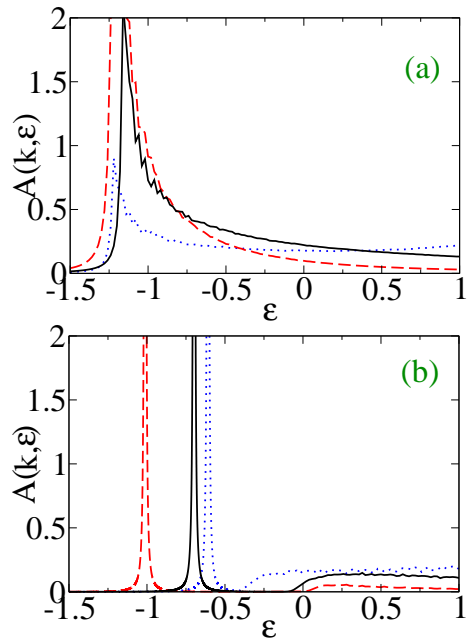

    \begin{center}
     \includegraphics[width=0.7\columnwidth,clip]{MD231A.eps}
     \includegraphics[width=0.7\columnwidth,clip]{MD300A.eps}
    \end{center}
\vspace{-15pt}
    \caption{(color online) The hole spectral functions at $t=0.5$, $t'=t''=0$, for $J_{\perp}=2.31$ $(a)$ and $J_{\perp}=3$ $(b)$. Here we show  ${\bf k}=(\frac{\pi}{2},\frac{\pi}{2})$ (black, solid), $(\pi,\pi)$ (red, dashed), and $(0,0)$ (blue, dotted). There is a significant distant incoherent part, the quasiparticle residues for $J_{\perp}=3$ are $Z_{(\frac{\pi}{2},\frac{\pi}{2})}=0.66$, $Z_{(\pi,\pi)}=0.88$,  $Z_{(0,0)}=0.32$. }
\label{0500spectra}
\end{figure}

\begin{figure}
    \begin{center}
     \includegraphics[width=0.85\columnwidth,clip]{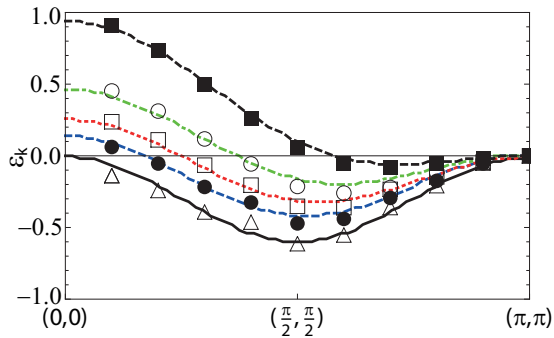}
    \end{center}
    \caption{(color online) The quasiparticle dispersion $\epsilon_k$ along the nodal direction. For a convenient comparison we choose $\epsilon_{\pi,\pi}=0$. Calculations are performed for $ t = 0.5J$, $t' = 0.0$, $t'' = 0.1J$ and for various values of the interlayer coupling $J_{\perp}$. Points show results of the dimer series expansion method and curves show results of SCBA. Here we take $J_{\perp} = 2.31J$ (bottom), $2.5J$, $2.7J$, $3J$ and $4J$ (top).}
    \label{fig:scbaseriest050tp00tpp010}
\end{figure}

\begin{figure}
    \begin{center}
     \includegraphics[width=0.85\columnwidth,clip]{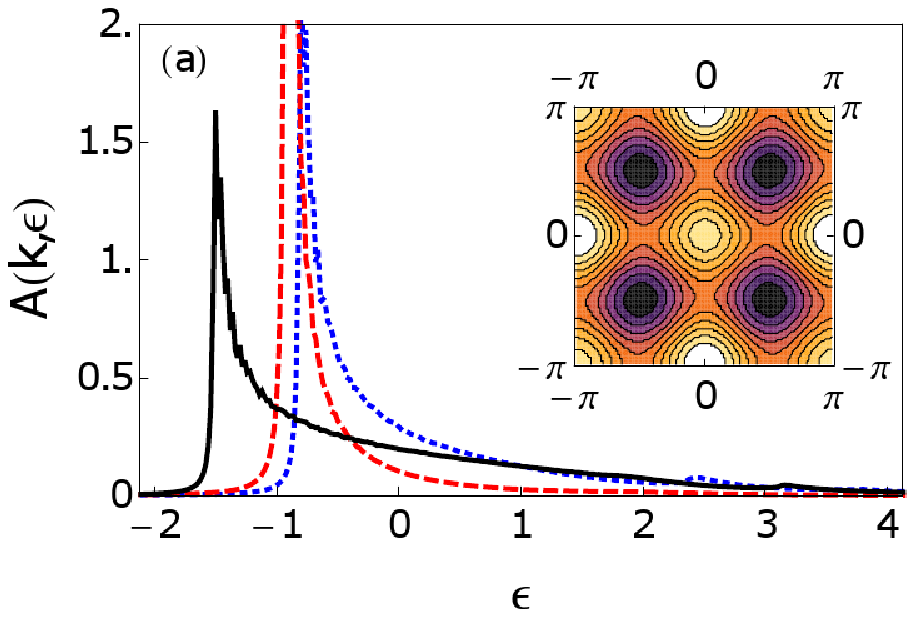}
     \includegraphics[width=0.85\columnwidth,clip]{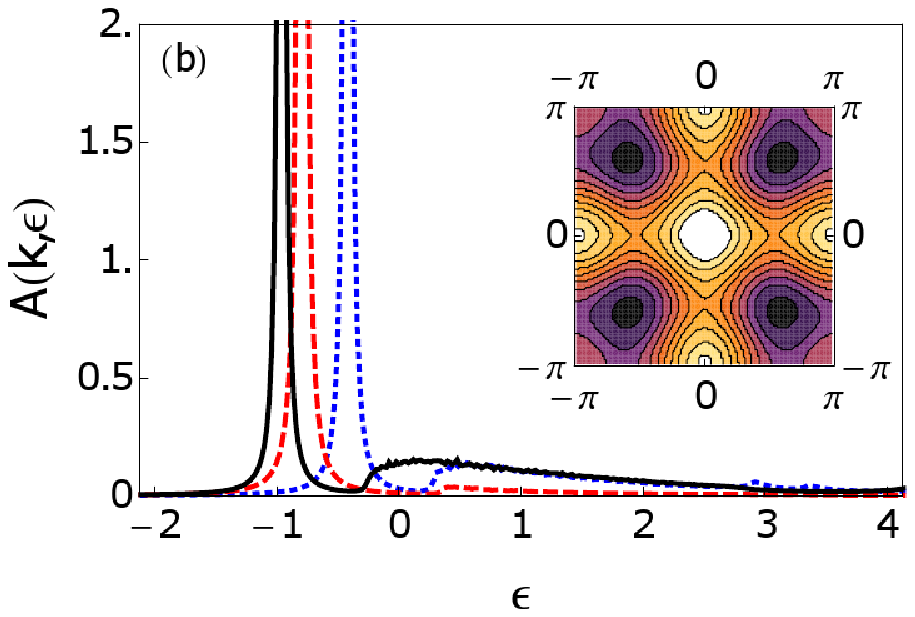}
     \includegraphics[width=0.85\columnwidth,clip]{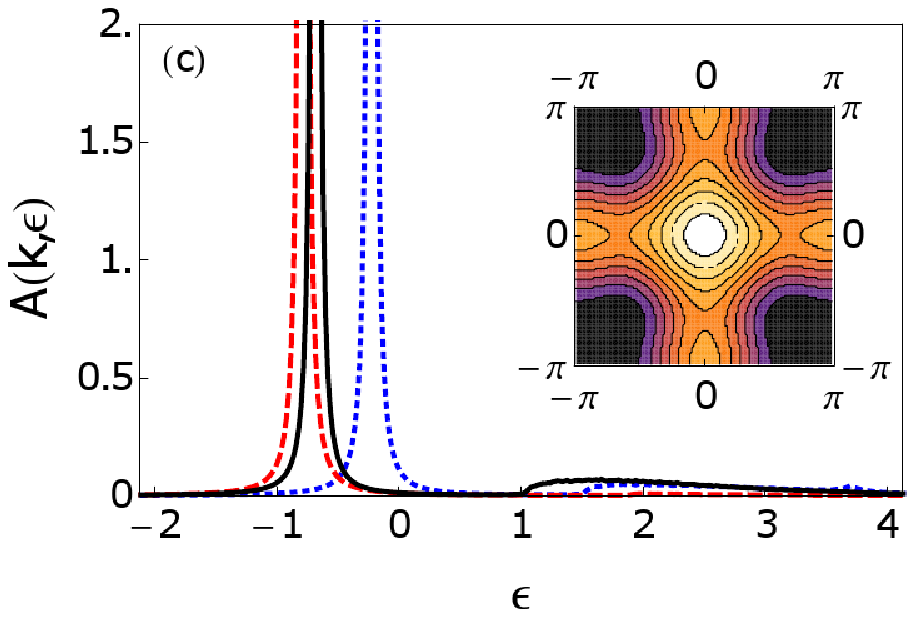}
    \end{center}
\vspace{-20pt}
    \caption{(color online) The hole spectral function at $t = 0.5$, $t' = 0.0$, $t'' = 0.1$ for  $J_{\perp} = 2.31$(a), $3.00$(b) and $4.00$(c).  Values of momentum are ${\bf k}=(\frac{\pi}{2},\frac{\pi}{2})$ (black, solid), $(\pi,\pi)$ (red, dashed), and $(\pi,0)$ (blue, dotted). The quasiparticle residues for $J_{\perp}=3$ are $Z_{(\frac{\pi}{2},\frac{\pi}{2})}=0.63$, $Z_{(\pi,\pi)}=0.90$,  $Z_{(0,0)}=0.32$. The insets in Fig.(a)-(c) show maps of the hole dispersion. The dark region is minimum of the dispersion.}    
\label{fig:spectralfermit050tp00tpp010}
\end{figure}

We start from the case of small hopping, $t=0.5$, $t'=t''=0$. At this small hopping we have two numerical methods, the SCBA and dimer series expansions. We can compare results and hence judge the accuracy of the methods. This is especially important since vertex corrections are neglected in the SCBA. We plot in Fig. \ref{0500spectra} the spectral functions at values $J_{\perp}=2.31$ and $J_{\perp}=3$. Note that $J_{\perp}=2.31$ is exactly the position of the QCP obtained from the mean field triplon analysis. We already pointed out that this position of the QCP is somewhat lower than that known from exact numerical calculations, $J_{\perp} = 2.52J$. Since in the SCBA we use triplon spectra from Section \ref{sec:bondoperatormft} we must refer to the consistent position of the QCP.

The spectra presented in Fig.\ref{0500spectra}(a) do not show any  quasiparticle peaks but instead only power cuts are observed. This behaviour of the mobile hole spectral function is similar to the Green's function of an immobile magnetic impurity at the QCP~\cite{Sushkov00, Vojta00}, implying that remarkably the hole mobility does not influence this behaviour.  The power cuts imply that the spin is distributed around the hole in a diverging cloud indicating SCS at the QCP (this will be discussed in more detail below).  On the other hand spectra in Fig. \ref{0500spectra}(b)  show quasiparticle peaks separated by the triplon gap $\Delta$ from the incoherent spectra. Figure \ref{0500spectra}(b) shows the dispersion minimum at  ${\bf k}=(\pi,\pi)$, while the cut position in Fig. \ref{0500spectra}(a) is essentially the same for all momenta. Hence, we conclude that the position of the LP for these parameters coincides with  that of the QCP. 

We also performed dimer series expansion calculations for both $t=0.5$, $t'=t''=0$ and $t = 0.5J$, $t' = 0.0$,  $t''=0.1$ and compared with the results of the SCBA calculations. Note that the series expansion method allows one to determine only  the quasiparticle dispersion. Naturally the method does not converge close to the QCP, as there are no quasiparticles there. However, at $J_{\perp}=3$ the method works well and agreement  between the SCBA and series is good, for example the SCBA band width is $\epsilon_{(0,0)}-\epsilon_{(\pi,\pi)}=0.40$ and the series band width is $0.41$ for the parameter set $t=0.5$, $t'=t''=0$.

In Fig. \ref{fig:scbaseriest050tp00tpp010} we show the comparison of the SCBA (lines) and dimer series expansions (points) for $t = 0.5J$, $t' = 0.0$,  $t''=0.1$. The series are well converged, even at the smallest value of  $J_{\perp}$. Hence the very good agreement between the two methods seen in Fig.  \ref{fig:scbaseriest050tp00tpp010} is strong evidence for the reliability of the SCBA approach for the present problem. In Fig. \ref{fig:scbaseriest050tp00tpp010} the dispersion minimum at the QCP is ${\bf k} = (\pi/2, \pi/2)$. At higher $J_{\perp}$ the minimum moves towards ${\bf k} = (\pi, \pi)$ reaching ${\bf k} = (\pi, \pi)$ at about $J_{\perp} \approx 4.5J$, this is the LP. This is confirmed by the spectral functions and FS maps shown in Fig. \ref{fig:spectralfermit050tp00tpp010}. Thus the LP in this case is well separated from the QCP deep in the magnetically disordered phase.

\begin{figure}
    \begin{center}
\includegraphics[width=0.85\columnwidth,clip]{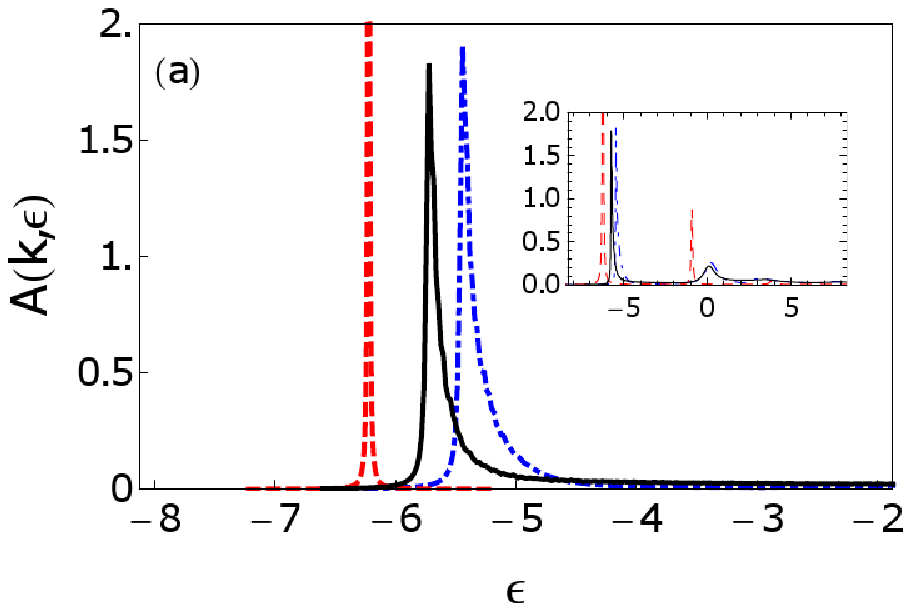}
\includegraphics[width=0.85\columnwidth,clip]{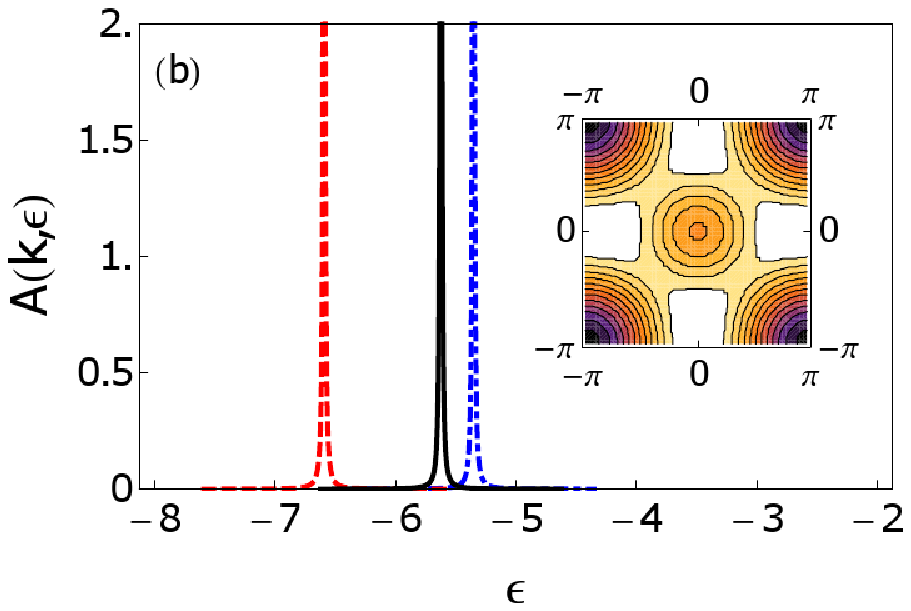}
    \end{center}
\vspace{-22pt}
    \caption{(color online) The hole spectral function at $t = 3.1$, $t'=t''= 0$, for $J_{\perp}=2.31$ $(a)$ and $J_{\perp}=3$ $(b)$. Values of the momentum are ${\bf k}=(\frac{\pi}{2},\frac{\pi}{2})$ (black, solid), $(\pi,\pi)$ (red, dashed), and $(\pi,0)$ (blue, dotted). The quasiparticle residues for $J_{\perp}=3$ are $Z_{(\frac{\pi}{2},\frac{\pi}{2})}=0.31$, $Z_{(\pi,\pi)}=0.80$,  $Z_{(\pi,0)}=0.35$. The inset in Fig.(a) shows spectral functions in the broader energy range and  the inset in Fig.(b) shows the map of the hole dispersion. The dark region is minimum of the dispersion.}
    \label{fig:spectralfermit310tp00tpp00}
\end{figure}

In the strong coupling limit $t = 3.1$ we rely on the SCBA since the series expansion does not converge in this strong coupling limit. Spectral functions for $t=3.1$, $t'=t''=0$ are shown in Fig.  \ref{fig:spectralfermit310tp00tpp00} for $J_{\perp}=2.31$ and $J_{\perp}=3.00$. In this case there is no LP in the disordered phase since the spectral functions show the bottom of the band is always at ${\bm k}=(\pi,\pi)$. Hence for these parameters the LP is inside the magnetically ordered phase in agreement with Ref.~\cite{Vojta99}. It is worth noting that at the band bottom there are well defined quasiparticles even at the QCP, indicating no SCS for this choice of parameters. The spectra at  ${\bm k}=(\frac{\pi}{2},\frac{\pi}{2})$ and ${\bm k}=(\pi,0)$  have cuts at the QCP, however, these are the high energy states which are irrelevant at small doping.

\begin{figure}
    \begin{center}
     \includegraphics[width=0.85\columnwidth,clip]{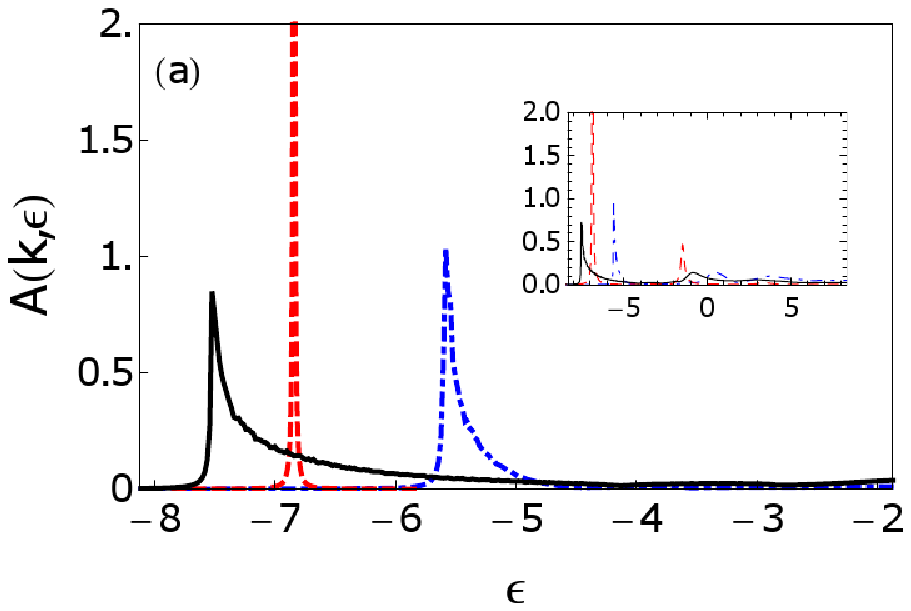}
     \includegraphics[width=0.85\columnwidth,clip]{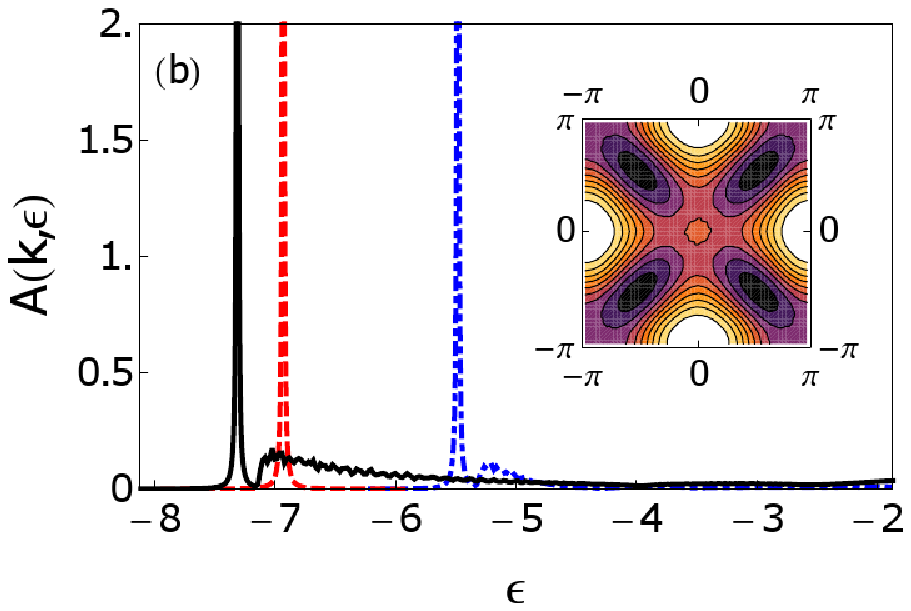}
     \includegraphics[width=0.85\columnwidth,clip]{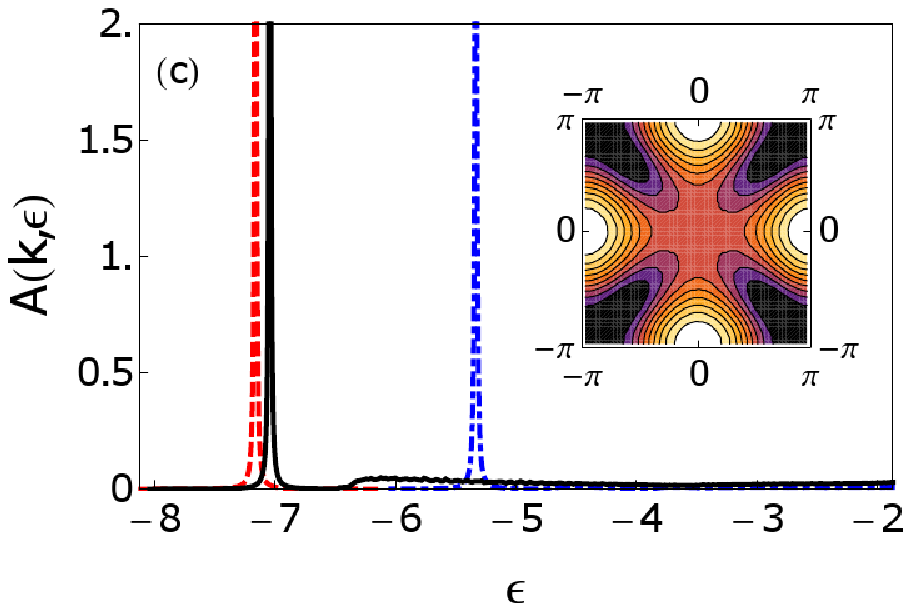}
     \includegraphics[width=0.85\columnwidth,clip]{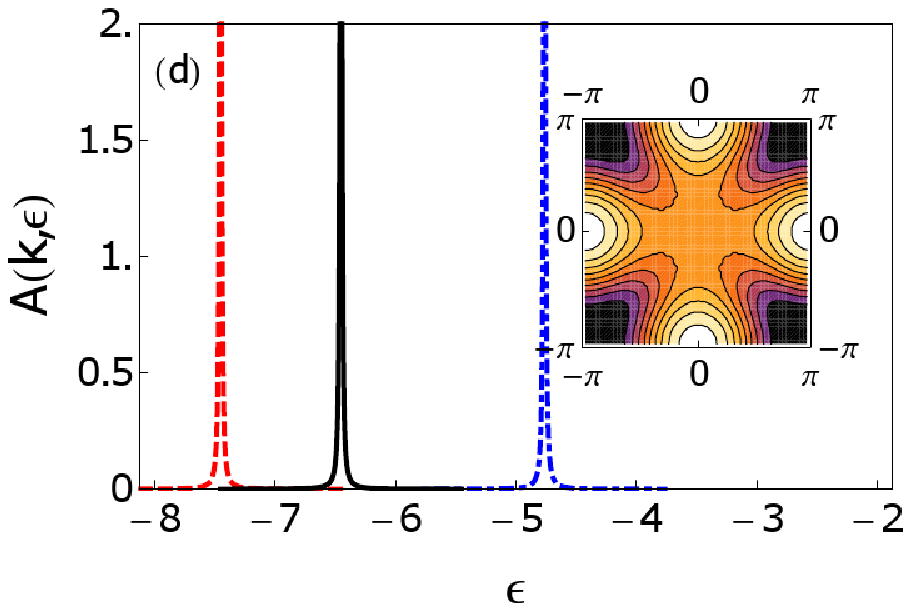}
    \end{center}
\vspace{-20pt}
    \caption{(color online) The hole spectral function at $t = 3.1$, $t' = -0.8$, $t'' = 0.7$ for  $J_{\perp} = 2.31$(a), $2.50$(b), $3.00$(c) and $4.00$(d).  Values of momentum are ${\bf k}=(\frac{\pi}{2},\frac{\pi}{2})$ (black, solid), $(\pi,\pi)$ (red, dashed), and $(\pi,0)$ (blue, dotted). The quasiparticle residues for $J_{\perp}=3$ are $Z_{(\frac{\pi}{2},\frac{\pi}{2})}=0.29$, $Z_{(\pi,\pi)}=0.77$,  $Z_{(\pi,0)}=0.20$; The inset in Fig.(a) shows spectral functions in the broader energy range and  the insets in Fig.(b)-(d) show  maps of the hole dispersion. The dark region is minimum of the dispersion.}    
\label{fig:spectralfermit310tp080tpp070}
\end{figure}

\begin{figure}
    \begin{center}
     \includegraphics[width=0.9\columnwidth,clip]{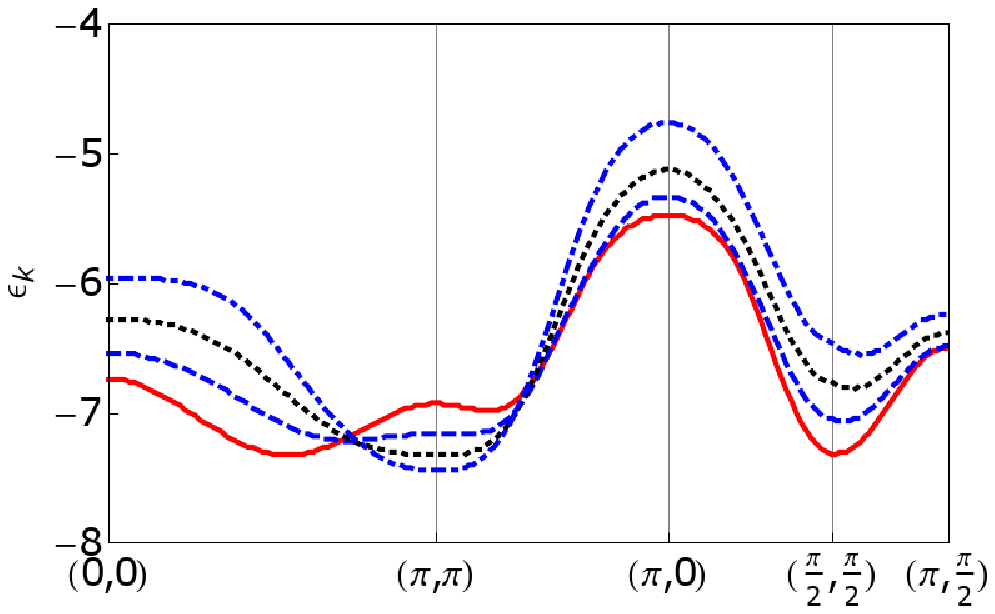}
     \includegraphics[width=0.9\columnwidth,clip]{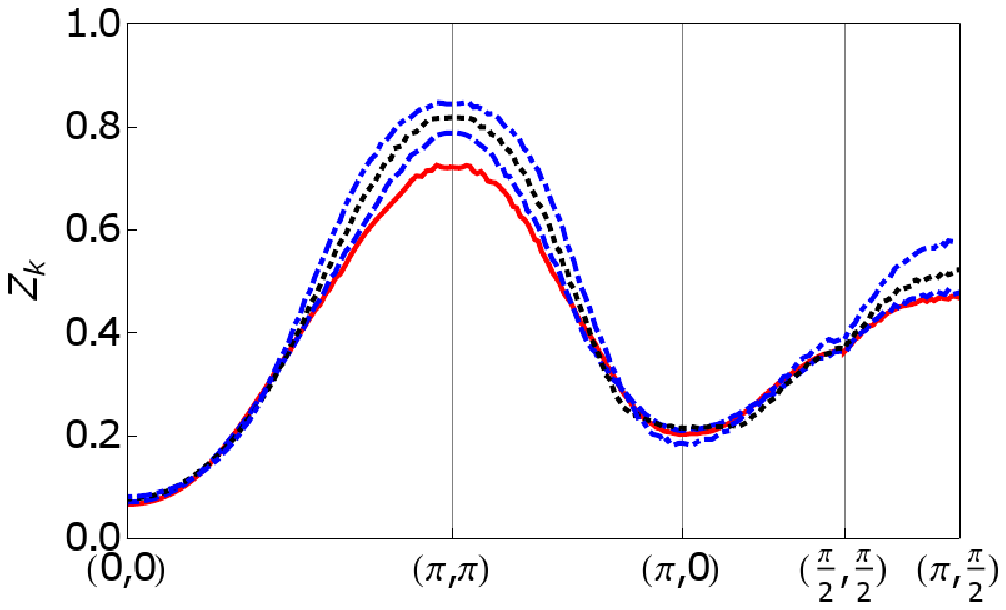}
    \end{center}
    \caption{(color online) Quasiparticle dispersion $\epsilon_k$ and residue $Z_k$ for different values of the interlayer coupling $J_{\perp}$. Calculations are performed with $t = 3.1J$, $t' = -0.8J$ and $t'' = 0.7J$. Here we considered $J_{\perp} = 2.5J$ (red, solid), $3J$ (blue, dashed), $3.5J$ (black, dotted) and $4J$ (blue, dot-dashed).} 
    \label{fig:dispQPresidt310tp080tpp070}
\end{figure}

The last and the most important set of parameters, $t = 3.1$, $t' = -0.8$, $t'' = 0.7$, roughly corresponds to the parameters of the cuprates \cite{Tokura90,Andersen95}. Although the current study does not intend to simulate the quantitative details of the cuprates, we choose the parameters relevant to the cuprates such that the interplay of various mechanisms can be compared on a similar energy scale. The spectral functions  are shown in  Fig. \ref{fig:spectralfermit310tp080tpp070} for several values of $J_{\perp}$. The dispersion maps shown in  the insets clearly demonstrate that the LP is located at $J_{\perp}\approx 3$ well within the magnetically disordered phase. In Fig. \ref{fig:dispQPresidt310tp080tpp070} we also present plots of the quasiparticle dispersion and quasiparticle residue, which also clearly demonstrate that the LP is located at $J_{\perp}\approx 3$.
This topological transition is caused by fully dynamic antiferromagnetic correlations in the absence of any static magnetic order. This demonstration of the possibility of the fully dynamic scenario is the first major conclusion of the present work.

It is important to ask how sufficiently small a doping is necessary for our analysis to be correct. In the present work we assume small doping, although it is possible to make a more quantitative estimate. It can be seen that practically we need $x < 0.1$ when the FS built on maps Fig \ref{fig:spectralfermit310tp080tpp070} (b) and Fig. \ref{fig:spectralfermit310tp080tpp070} (c) are topologically different. Another important question is why have the longer range hoppings have qualitatively changed the situation? The reason is this: at $t' = t'' = 0$ the LP is in the ordered phase but already close to the QCP. A hopping $t'' > 0$ pushes the bare dispersion \eqref{eq:fullholedispersion} at the nodal point ${\bm k}=(\frac{\pi}{2},\frac{\pi}{2})$ down helping magnetic fluctuations to form a small pocket. A pretty small positive $t''$ is sufficient to shift the LP to the disordered phase. The role of $t'$ is less important. The shift of the LP is due to the tuning of the longer range hoppings. The "tuning" has been performed by nature in the cuprates where the qualitative importance of $t'$, $t''$ is well known. These parameters give asymmetry between the hole and the electron doping. Holes go to the nodal points while electrons go to the antinodal ones resulting in dramatically different Fermi surfaces and magnetic properties. We follow nature and rely on the same mechanism.

\section{Spin-Charge Separation}

\label{sec:scs}

The second major conclusion of the present work is related to the first one and concerns SCS at the QCP. Looking at Fig.\ref{fig:spectralfermit310tp080tpp070}(a), which is at the QCP, we observe that the lowest energy spectral function corresponding to ${\bf k}=(\frac{\pi}{2}, \frac{\pi}{2})$, does not have a pole, but only a cut. According to previous studies for the case of an immobile impurity, the presence of a cut indicates that the spin density is distributed in a power cloud around the hole \cite{Sushkov00, Vojta00}. Because the Green's function in the present case is similar to that of the immobile impurity we directly project the results of Refs. \cite{Sushkov00, Vojta00} to understand SCS at the QCP in the present case. When approaching the QCP from the magnetically disordered phase the quasiparticle residue approaches zero, $Z \propto \Delta^z$, $z\approx 0.4$, as the triplon gap $\Delta$ approaches zero. The fraction of spin localized at  hole  goes to zero $\propto Z$. The rest of spin is distributed around the hole over a disk of radius $R\propto 1/\Delta$. At $r \ll R$ the spin density is  $\propto 1/r^{\alpha}$, $1 < \alpha < 1.5$. Therefore the average radius of the spin cloud $\langle r\rangle \sim R$ diverges at the QCP, indicating SCS.  As one moves away from the QCP into the magnetically disordered phase, the quasiparticle peak appears, but the triplon gap is still significant as long as the system is close to the QCP (as can be seen in Fig. \ref{fig:spectralfermit310tp080tpp070}(b)). This indicates a significant amount of spin is still relatively delocalized from the hole, and a rather smooth crossover to the spin charge recombined region.  On the other side of the QCP, deep inside the AF phase, the hole interaction with a magnetic field is described by pseudospin \cite{Braz89, Ramaz08} and this interaction implies a partial SCS \cite{Milstein08}. Evolution of the spin cloud when approaching the QCP {\it from the AF phase} is not clear at present. 

Because of the diverging size of the magnon cloud the hole effective mass also diverges at the QCP.  Drawing analogy with the cuprates we note that the effective mass measured in MQO \cite{Singleton10} diverges on approaching the QCP as identified by neutron scattering studies \cite{Stock08, Hinkov08, Haug10}.

\section{Conclusions}

\label{sec:conclusions}

We provide a detailed theory for the results presented in Ref. \onlinecite{Holt12} where we considered the bilayer $t-t^{\prime}-t^{\prime\prime}-J$ model with strong interlayer coupling $J_{\perp}$. In our analysis we assumed that the hole doping is low and does not influence the magnetic dynamics and hence fill rigid bands formed by quantum magnetic fluctuations. We have argued that our results should be valid for doping $x < 0.10$. Therefore, the magnetic dynamics are driven only by the interlayer coupling $J_{\perp}$ with a magnetic QCP at $J_{\perp} \approx 2.5J$ separating the magnetically ordered and magnetically disordered phases.  Comparison between the hole dynamics described by the SCBA and series expansion methods are made explicitly, which shows excellent agreement. 

At  $J_{\perp} \to \infty$ the hole Fermi surface is connected and centered at  ${\bf k}=(\pi,\pi)$. We found that at a certain $J_{\perp}^{LP}$  the AF correlations reconstruct the connected Fermi surface into four seperate pockets centered close to  ${\bf k}=(\pm \frac{\pi}{2},\pm \frac{\pi}{2})$.  We have demonstrated that the LP where the topology of the Fermi surface is changed can be located within the magnetically disordered phase. The LP is driven by purely dynamic antiferromagnetic correlations in absence of any  static magnetic order.  The precise position of the LP at which this topological transition occurs depends on hole hopping integrals $t^{'}$ and $t^{''}$. This is because the reason behind the shifting of the hole pockets is due to magnetic fluctuations that diminish the nearest neighbor hopping $t$. The $t^{'}$ and $t^{''}$ then have the tendency to create hole pockets near ${\bf k}=(\pi/2,\pi,/2)$. The precise position of LP depends on the balance of these two mechanisms. In fact, for certain values of $t^{'}$ and $t^{''}$ one may not see such transition if they are too weak to overcome the residual $t$. Nevertheless, for the parameters that are closest to the real cuprates, one sees such a transition at $J_{\perp}/J\approx 3$, as indicated in Fig. \ref{fig:spingapvelocity}. 

We have also demonstrated that if the LP is located in the magnetically disordered phase then, on approaching the magnetic QCP, the hole spin and charge separate. The separation scale is equal to the magnetic correlation length which diverges at the QCP.

We are not aware of any 2D materials which have a low concentration of charge carriers and a magnetic QCP driven by a separation parameter. Perhaps such materials will be synthesised in the future and/or the model can be realised in cold atoms. However, the most important outcome of the analysis is the demonstration of the principal possibility of the dynamic magnetic fluctuation driven topological transition and spin-charge separation.  We anticipate that this calculation, although it does not describe the cuprate physics directly, indicates several important ingredients in the underdoped cuprates that are very likely to be the origin of FS reconstruction and phenemena related to SCS.

\section{Acknowledgements}

Numerical calculations performed in this work were done at the National Computational Infrastructure at the ANU, under project u66.

\end{document}